\documentclass[12pt, a4paper]{article}

\usepackage[dvips,letterpaper]{geometry}
\usepackage{graphicx}
\usepackage{times}
\usepackage{setspace}
\usepackage{amsthm, amsmath, natbib, amssymb, amsfonts,xparse}
\usepackage{mathtools, here}
\usepackage{breakcites}
\usepackage{bm}
\usepackage{subcaption}
\usepackage{multirow}
\usepackage{url}
\usepackage{verbatim}
\usepackage{blkarray}
\usepackage{natbib}
\usepackage{color}
\usepackage{booktabs}
\usepackage{cool}
\usepackage{physics}
\usepackage{lineno}
\usepackage{listings}
\usepackage{tikz}

\usetikzlibrary{matrix,shapes,arrows,positioning}

\definecolor{codegreen}{rgb}{0,0.6,0}
\definecolor{codegray}{rgb}{0.5,0.5,0.5}
\definecolor{codepurple}{rgb}{0.58,0,0.82}
\definecolor{backcolour}{rgb}{0.95,0.95,0.92}

\lstdefinestyle{mystyle}{
    backgroundcolor=\color{backcolour},
    commentstyle=\color{codegreen},
    keywordstyle=\color{magenta},
    numberstyle=\tiny\color{codegray},
    stringstyle=\color{codepurple},
    basicstyle=\footnotesize,
    breakatwhitespace=false,
    breaklines=true,
    captionpos=b,
    keepspaces=true,
    numbers=left,
    numbersep=5pt,
    showspaces=false,
    showstringspaces=false,
    showtabs=true,
    tabsize=2}

\input cyracc.def

\usepackage[breaklinks=true,pdfencoding=auto]{hyperref}

\makeatletter
\newcommand\fake@math{}
\def\fake@math#1\){[math]}
\makeatother

\newtheorem{Corollary}{Corollary}[section]

\newtheorem{Proposition}{Proposition}[section]

\newtheorem{Proof}{Proof}[section]

\providecommand{\keywords}[1]
{
  \small	
  \textbf{\textit{Keywords---}} #1
}

\title{Quantile Regression for positive data using a general class of distributions}
\vspace{-0.25cm}
\author{
\textbf{Diego I. Gallardo}$^{1}$ \,\,\textbf{and} \,\,  \textbf{Manoel Santos-Neto}$^{2}$
\\
{\small $^{1}$Departamento de Matem\'atica, Facultad de Ingenier\'ia, Universidad de Atacama, Copiap\'o, Chile}\\
{\small $^{2}$Departamento de Estat\'istica, Universidade Federal de Campina Grande, Campina Grande, Brasil}\\ [-0.15cm]
}

\date{}

\begin{document}

\maketitle

\begin{abstract}
This paper presents a general
class of quantile regression 
models for positive continuous 
data. In this class of models we
consider that the response variable
has a IRON distribution.
We provide inference and diagnostic tools for this class of models. An R package, called IRON, was implemented. This package provides estimation and inference for the parameters and tools useful to check the fit of models. The methods are also illustrated with an application to modeling household income in Chile.
\end{abstract}


\section{Introduction}
The family of distributions proposed by \cite{Lehmann1953} may be characterized by its cumulative distribution function (c.d.f) given by
\[
G(z\mid {\bm \theta}, \alpha) = \Pr(Z \leq z \mid {\bm \theta}, \alpha) = [F(z\mid \vb*{\theta})]^\alpha \qc z \in \real,  \, \alpha \in \real_{>0},\,
\vb*{\theta} \subset \real^p \,(\text{where}\; p=\textrm{dim}(\vb*{ \theta})), 
\]
and is called of standard $\alpha$-exponentiated distribution and use the notation $Z \sim \textbf{EXP}f_{\bm \theta}(\alpha)$. 
As $\alpha=1$ include as particular case the basal model, many distributions has been extended considering this method. To name a few, we have the exponentiated-Weibull \citep{Mudholkar1995}, the exponentiated-exponential \citep{GuptaKundu2001}, the exponentiated log-normal \citep{Kakde06}, the exponentiated gamma \citep{Nadarajah07} and the power piecewise exponential \citep{Gomez2018} models, among others.
Particularly, \cite{Martinez-Florez2014} developed the standard $\alpha$-exponentiated Birnbaum-Saunders (\textbf{EXPBS}$f$) distribution, which c.d.f is given by
\begin{equation*}
G(t|\lambda,\beta,\alpha)=[F(a_t)]^\alpha\qc t,\lambda,\beta,\alpha \in \real_{>0},
\end{equation*}
where $F(\cdot)$ is a c.d.f related to a distribution for a random variable with support in $\real$ and $a_t = \lambda^{-1}\left(\sqrt{t/\beta}-\sqrt{\beta/t}\right)$. Besides that, the probability density function (p.d.f) is
\begin{equation*}\label{eq:3}
 g(t|\lambda,\beta,\alpha) = \alpha f(a_t)[F(a_t)]^{\alpha-1}\vdot \frac{t^{-3/2}(t+\beta)}{2\lambda\sqrt{\beta}}\qc t,\lambda,\beta,\alpha \in \real_{>0},
\end{equation*}
where $f(\cdot) = \dd F(\cdot)$. The Birnbaum-Saunders (\textbf{BS}) distribution~\citep[][]{Birnbaum1969} corresponds to the case where $\alpha=1$ and $F=\Phi$, the c.d.f of the standard normal distribution. 
Without loss of generality and only for simplicity purpose, we consider that $g_t = g(t|\lambda,\beta,\alpha)$.  
For any $\tau \in (0,1)$ 
\begin{equation}\label{eq:qf}
 F^{-1}(\tau) = \inf\left\{t| F(t) \geq \tau \right\},
\end{equation}
is called the $\tau$th quantile of $T$. To ensure that, $F^{-1}(1/2) = 0$, we will consider elliptical distributions about the origin. This result will be important for the formulation of the model proposed in this work. 

Let $Y$ be a random variable with elliptical distribution with location parameter $\mu \in \real$, dispersion parameter $\sigma \in \real_{>0}$, symmetric kernel $k(\cdot)$ and p.d.f given by
\begin{equation}\label{eq:5}
 f(y | \mu, \sigma, k  ) = \frac{1}{\sigma} \vdot k\left[\left(\frac{y-\mu}{\sigma}\right)^2\right]\qc y \in \real,
\end{equation}
where the function $k: \real \to \real_{>0}$ is such that $\int\limits_{0}^\infty k(u)\dd u$ and $\int\limits_{0}^{\infty}u^{-1/2}k(u)\dd u$ are finite. We have, when they exist, that $\textbf{E}(Y) = \mu$ and $\textbf{Var}(Y) = \phi \sigma$, where $\phi>0$ is a constant that may be obtained from the expected values of the radial variable or from the derivative of the characteristic. For example, for the Student's $t$ distribution with $\xi$ degrees of freedom one has $\phi = \xi/(\xi-2)$ for $\xi>2$. If $\mu=0$ and $\sigma = 1$ (the standard case) we can be rewriting \eqref{eq:5} as
\begin{equation}\label{eq:sym}
  f(y| 0, 1, k  ) =  k\left(y^2\right)\qc y \in \real,
\end{equation}
and we will denote $Y \sim EL(0, 1, k)$. For some expressions of $k(x)$ and $F(x)$ for several elliptical distributions, see  Table~\ref{tab:1}.
\begin{table}[H]
 \caption{Expressions of $k(x)$ and $F(x)$ for some standardized elliptical distributions.}
 \centering
  \begin{tabular}{ccc} \toprule
   Distribution & $k(x)$ & $F(x)$ \\ \midrule
    Normal & $\frac{1}{\sqrt{2\pi}}\exp(x/2)$ & $\Phi(x)$  \\
    Student' $t$&  $\frac{\xi^{\xi/2}}{\Beta{1/2,\xi/2}}(\xi+x^2)^{-\frac{\xi+1}{2}} $& $\frac{1}{2} + \frac{x\, \Hypergeometric{2}{1}{1/2,1/2(\xi+1)}{3/2}{-x^2/\xi} }{\sqrt{\xi}\,\Beta{\frac{\xi}{2},\frac{1}{2}}}$\\
    Logistic& $\frac{\exp(x)}{[1+\exp(x)]^2}$ & $\frac{\exp(x)}{1+\exp(x)}$ \\
    Exponential Power& $\frac{\kappa}{2\Gamma(\kappa)}\exp\left(|x|^\kappa\right)$ & $\frac{1}{2}\left[1 + \frac{\textrm{sign}(x)}{\Gamma(1/\kappa)}\gamma(1/\kappa,|x|^\kappa)  \right]$ \\
     Cauchy & $\frac{1}{\pi(1 + x^2)}$   & $\frac{1}{\pi}\arctan(x) + \frac{1}{2}$  \\
   \bottomrule
     \multicolumn{3}{l}{\footnotesize $\Beta{a,b}$ is the beta function and $\Gamma(a)$ is a gamma function.}\\
   \multicolumn{3}{l}{\footnotesize $\Hypergeometric{2}{1}{a}{b}{x}$ is the hypergeometric function.}\\
    \multicolumn{3}{l}{\footnotesize $\Phi(x) = \frac{1}{2}\Erf{ \frac{x}{\sqrt{2}} }$ where $\Erf{\cdot}$ is the error function.}\\
   \multicolumn{3}{l}{\footnotesize $\gamma(s,x) = \int\limits_0^x t^{s-1}\exp(-t)\textrm{d}t$ is the lower incomplete gamma function.}
   \end{tabular}
 \label{tab:1}
 \end{table}

Considering the density function given in \eqref{eq:sym} we have the following log density
\begin{align}\label{eq:ll}
\log(g_t) =&  \underbrace{\log(k\left( {a^*_t}^2 \right)) + (\alpha -1)\log(\Psi(a_t))}_{\text{specific term}(\log(g_t^s))}  \nonumber \\ 
&+ \underbrace{\log(\alpha) - \frac{3}{2}\log(t) + \log(t+\beta) - \log(2\lambda) - \frac{1}{2}\log(\beta)}_{\text{common term}(\log(g_t^c)) }.
\end{align}
where $\Psi(a_t) = \int\limits_{-\infty}^{a_t} k\left( u^2\right) \dd u  $ is the c.d.f of the $EL(0, 1,k)$. Note that, the log density in \eqref{eq:ll}
is divided in a specific term that depend of the kernel selected and a 
common term associated to Birnbaum-Saunders distribution.

\cite{Martinez-Florez2014} discussed the flexibility of the \textbf{EXPBS}$f$ and some properties, such as moments and quantiles. However, such expressions are complicated and depend on each specific selection for $F(\cdot)$. Morevover, none of the parameters can be interpreted as a useful measure from the population (such as mean, mode, among others), except in the usual case $\alpha=1$, where $\beta$ is the median of the population.

To counter this problem, henceforth we focuses on $F(\cdot)$ related to standardized elliptical distributions with support in $\real$ (see, Table~\ref{tab:1}). This variation of the \textbf{EXPBS}$f$ family, which generalizes the \cite{DIAZGARCIA2005445} family, can be called of
\textbf{IRON} distribution due to the possibility of considering the heavy tails distributions. Additionally, in order to distinguish among the different models for $F$, we use the notation \textbf{IRON-N}, \textbf{IRON-t}, \textbf{IRON-L}, \textbf{IRON-EP} and \textbf{IRON-C} for the case of the normal, Student't, logistic, power exponential and Cauchy models, respectively.

Quantile regression has had intense research activity in recent years for parametric models. See for instance, \cite{Galarza2017}, \cite{Gallardo2020} and \cite{Mazucheli2020}. However, up to this moment we only find the work of \cite{https://doi.org/10.1002/asmb.2556} related to the application of a \textbf{BS}-type distribution in this context. For those reasons,
the aim of this paper is to provide a quantile regression model for positive continuous response variables based on a general class of \textbf{BS}-type distributions. More formally,  we will consider $\alpha_\tau=-\log(\tau)/\log(2)$ fixed and we have, as a result 
\begin{equation}
G(\beta|\lambda,\beta,\alpha_\tau)=[F(0)]^{\alpha_\tau}=(1/2)^{\alpha_\tau} = \tau, \label{eq.tau}
\end{equation}
i.e. fixing $\alpha = \alpha_\tau$, $\beta$ denotes the $\tau$th quantile of the distribution as defined in \eqref{eq:qf} for any elliptical c.d.f $F$. For $\alpha_\tau$, we denote the model as \textbf{RIRON}$_\tau $-F$(\beta,\lambda)$. Therefore, we define a rich class to perform quantile regression for positive data (not only for median regression) based on a class of \textbf{BS} type distributions.
This article is divided as follows. Section~\ref{sec:2} presents a new proposal for quantile regression in a \textbf{BS}-type model. Section \ref{sec:3} discuss the diagnostic, residual analysis and some computational aspects for the model. Two simulation studies are reported in Section \ref{sec:4}. Section \ref{sec:5} presents a real data application related to the last study available (from 2016) for the chilean house-hold income. Finally, Section \ref{sec:6} presents final remarks related to our proposal.


\section{A quantile regression model based on the BS distribution}\label{sec:2}

\cite{https://doi.org/10.1002/asmb.2556} discussed a version of the \textbf{BS} model parametrized in terms of the $\tau$th quantile. This parametrization corresponds to take $Q_\tau=(\beta/4)(\lambda z_\tau+\sqrt{\lambda^2 z_\tau^2+4})^2$, where $z_\tau$ denotes the $\tau\times 100$th quantile of the standard normal model. We denote this model as \textbf{RBSQ}$_\tau(Q_\tau,\lambda)$. In this case, $Q_\tau$ is the $\tau$th quantile of the \textbf{RBSQ} model. The authors performed a regression analysis considering
\begin{equation}
\label{cs0}
\vb{link}(Q_{\tau i}) = \eta_i(\tau)= \vb{x}^\top_i\vb*{\psi}(\tau)\qc
(i = 1,\ldots, n),
\end{equation}
where $\vb*{\psi}(\tau) = (\psi_1(\tau), \ldots, \psi_p(\tau))^\top$ 
is vector of unknown regression coefficients,
$\vb*{\psi}(\tau) \in \mathbb{R}^p$, with $p< n$,
$\eta_{i}(\tau)$ is the linear predictor, and $\vb{x}_i = (x_{i1}, \ldots, x_{ip})^\top$ is observations on $p$ known regressors, for $i = 1, \ldots, n$. Furthermore, the authors assume that the matrix $\vb{X} = (\vb{x}_1, \ldots, \vb{x}_n)^\top$ have $\rank p$. Finally, $\vb{link}(\cdot)$ is strictly monotonic, invertible and a twice differentiable link function. In the fo\-llo\-wing proposition we prove that the \textbf{RBSQ}$_\tau$ model with any $\tau \in (0,1)$ and the regression structure defined in \eqref{cs0} define the same p.d.f.

\begin{Proposition}\label{prop1}
Let $Y_1 \sim \textbf{RBSQ}_\tau(Q_{\tau},\lambda)$ and $Y_2 \sim \textbf{RBSQ}_{\tau^\star}(R_{\tau^{\star}},\xi)$, where $Q_\tau=\vb{x}^\top\vb*{\psi}(\tau)$ and 
$R_{\tau^\star}=\vb{x}^\top\vb*{\varphi}(\tau^\star)$ and without loss of generality $0<\tau<\tau^\star<1$. If $\vb{x}$ includes an intercept term, then there is a one-to-one transformation from $(\vb*{\psi}(\tau)^\top,\lambda)$ to $(\vb*{\varphi}(\tau^\star)^\top,\xi)$, i.e., $Y_1$ and $Y_2$ are equal in distribution. 
\end{Proposition}

\begin{Proof}
Exploring the relation between the \textbf{BS} and the \textbf{RBSQ} distributions, for $Y_1$ we have that $Q_\tau=\beta \rho_\tau(\lambda)$ and $\alpha=\lambda$
and for $Y_2$ we have that $R_{\tau^\star}=\beta \rho_{\tau^\star}(\xi)$ and $\alpha=\xi$, where $\rho_\tau(u)=0.25\times (u z_{\tau}+\sqrt{u^2z_\tau^2+4} )^2$. Therefore, $\lambda=\xi$ and $Q_\tau=R_{\tau^\star} \times \rho_\tau(\lambda) / \rho_{\tau^\star}(\xi)$. The last equations implies that
\[
\vb{link}^{-1}\left(\vb{x}^\top\vb*{\psi}(\tau)\right)=\vb{link}^{-1}\left(\vb{x}^\top\vb*{\varphi}(\tau^\star)\right)\times \rho_\tau(\lambda) / \rho_{\tau^\star}(\xi).
\]
As $\vb{x}^\top$ include an intercept term we can write $\vb{x}=(1,\vb{x}^{*\top})^\top$, $\vb*{\psi}(\tau)=(\psi_0(\tau),\vb*{\psi}^*(\tau))$ and $\vb*{\varphi}(\tau)=(\varphi_0(\tau^\star),\vb*{\varphi}^*(\tau^\star))$ and then
\[
\vb{link}^{-1}\left(\psi_0(\tau)+\vb{x}^{*\top}\vb*{\psi}(\tau)\right)=\vb{link}^{-1}\left(\varphi_0(\tau^\star)+\vb{x}^{*\top}\vb*{\varphi}^*(\tau^\star)\right)\times \rho_\tau(\lambda) / \rho_{\tau^\star}(\xi).
\]
As $\vb{link}(\cdot)$ is strictly monotonic and invertible, we obtain the following relations for each choice of the link function.
\begin{itemize}
    \item \textbf{log}: $\psi_0(\tau)=\varphi_0(\tau^\star)+\log \rho_\tau(\lambda)-\log \rho_{\tau^\star}(\lambda)$ and $\vb*{\psi}(\tau)=\vb*{\varphi}(\tau^\star)$. 
\item \textbf{identity}: $\psi_0(\tau)=\varphi_0(\tau^\star)\times \rho_\tau(\lambda) / \rho_{\tau^\star}(\lambda)$ and $\vb*{\psi}(\tau)=\vb*{\varphi}(\tau^\star)\times \rho_\tau(\lambda) / \rho_{\tau^\star}(\lambda)$.
\item \textbf{squared root}: $\psi_0(\tau)=\varphi_0(\tau^\star)\times \sqrt{\rho_\tau(\lambda) / \rho_{\tau^\star}(\lambda)}$ and $\vb*{\psi}(\tau)=\vb*{\varphi}(\tau^\star)\times \sqrt{\rho_\tau(\lambda) / \rho_{\tau^\star}(\lambda)}$.
\end{itemize}
Note that in all the cases there is a one-to-one mapping from $(\vb*{\psi}(\tau)^\top,\lambda)$ to $(\vb*{\varphi}(\tau^\star)^\top,\xi)$, where follows the result. $\hfill\square$.
\end{Proof}

\begin{Corollary}
The \textbf{RBSQ}$_\tau$ model with a regression structure for $Q_\tau$ and $\lambda$ constant provides the same log-likelihood function (and then, the same criteria based on its such as AIC and BIC) for any $\tau \in (0,1)$. 
\end{Corollary}
\begin{Proof}
It is a direct consequence of Proposition \ref{prop1}. $\hfill\square$.
\end{Proof}

Provided the limitation of the \textbf{RBSQ}$_\tau$ model in terms of modelling, we propose to study the \textbf{EXPBS}$f$ model using the property in equation (\ref{eq.tau}) as a concurrent model for positive data in a quantile regression model context.
Considering $\alpha_\tau$ fixed, we suppose the $\tau$th quantile satisfies the following functional relation
\begin{equation*}\label{cs1}
\vb{link}(\beta_{\tau i}) = \eta_{i}(\tau) = \vb{x}^\top_i\vb*{\gamma}(\tau)\qc 
(i = 1,\ldots, n),
\end{equation*}
where $\vb*{\gamma}(\tau) = (\gamma_1(\tau), \ldots, \gamma_p(\tau))^\top$ 
is vector of unknown regression coefficients,
$\vb*{\gamma}(\tau) \in \mathbb{R}^p$, with $p< n$ and we assume the same assumptions for the matrix $\vb{X}$ and the function $\vb{link}(\cdot)$ mentioned previously. 
For this model, we have the following proposition.
\begin{Proposition}\label{prop2}
Let $Y_1 \sim $\textbf{RIRON}$_\tau $-F$(\beta_\tau,\lambda)$ and $Y_2 \sim $\textbf{RIRON}$_{\tau^\star}$-F$(\zeta_{\tau^\star},\xi)$, where $\beta_\tau=\vb{x}^\top\vb*{\psi}(\tau)$ and
$\zeta_{\tau^\star}=\vb{x}^\top\vb*{\varphi}(\tau^\star)$ and without loss of generality $0<\tau<\tau^\star<1$. Then, $Y_1$ and $Y_2$ are not equal in distribution. 
\end{Proposition}
\begin{Proof}
If $Y_1$ and $Y_2$ are equal in distribution, their c.d.f satisfy the following equality
\[
\left[F\left(\frac{1}{\lambda}\left(\sqrt{\frac{t}{\beta_\tau}}-\sqrt{\frac{\beta_\tau}{t}}\right)\right)\right]^{\alpha_\tau}=\left[F\left(\frac{1}{\xi}\left(\sqrt{\frac{t}{\zeta_{\tau^\star}}}-\sqrt{\frac{\zeta_{\tau^\star}}{t}}\right)\right)\right]^{\alpha_{\tau^\star}}, \quad \forall t>0.
\]
For the models considered in Table \ref{tab:1}, such equation is valid only if $\lambda=\xi$, $\beta_\tau=\zeta_{\tau^\star}$ and $\alpha_\tau = \alpha_{\tau^\star}$. The last one is verified only if $\tau=\tau^\star$, producing a contradiction. Therefore, $Y_1$ and $Y_2$ are not equal in distribution.
\end{Proof}

Henceforth, to simplify the notation we remove $\tau$ in the parameters. The logarithm of the likelihood function for the parameters vector $\vb*{\theta} = (\vb*{\gamma},\lambda, \sigma)^\top$ considering a random sample of $n$ observations is given by
  \begin{equation}\label{eq:loglik}
  \ell(\vb*{\theta}|\vb{y}) =  \sum_{i=1}^n \log(g_{t_i}^s)  + \sum_{i=1}^n \log(g_{t_i}^c),
  \end{equation}
 where $\log(g_t^s)$ and $\log(g_t^c)$ are defined in \eqref{eq:ll} with $\alpha = \alpha_\tau$. 
 
%
 %

\section{Residual and Diagnostic}\label{sec:3}

In this Section we present the generalized Cook’s distance (GSD) in order to detect potential influent observations.
We also present a kind of residual to discuss if the model is appropriated.
Additionally, we also discuss some computational aspects of the model.

\subsection{Generalized Cook's distance}
 
In this Section we use a generalization of the Cook's distance~\citep{10.2307/1268249, https://doi.org/10.1111/j.2517-6161.1986.tb01398.x}
with the objective of assess the influence of individual observations 
on the predicted conditional quantile of the response variable. In case
of our proposed model, such generalization is defined as
\[
\textrm{GCD}_i(\bm{\theta})=\frac{1}{q}\bigg[\big(\widehat{\bm{\theta}}-\widehat{\bm{\theta}}_{(i)}\big)^\top \widehat{\bm{\Sigma}}_{\widehat{\theta}}^{-1}\big(\widehat{\bm{\theta}}- \widehat{\bm{\theta}}_{(i)} \big)\bigg], \quad i = 1,\ldots,n,
\]
where $\bm{\theta} = (\bm \gamma^\top, \lambda, \xi)$, $q = \textrm{dim}(\bm{\theta})$, $\widehat{\bm{\Sigma}}_{^{\widehat{\theta}}}$ is an estimate of the variance-covariance matrix of $\widehat{\bm{\theta}}$, and $\widehat{\bm{\theta}}_{^{(i)}}$ is the MLE of $\widehat{\bm{\theta}}$ without considering the case $i$.  If the interest is just on the $p\times 1$ vector of regression coefficients, $\bm \gamma = (\gamma_1, \ldots, \gamma_{p})^\top$, then 
$$
\textrm{GCD}_i({\bm \gamma})=\frac{1}{p}\bigg[\big(\widehat{\bm{\gamma}}-\widehat{\bm{\gamma}}_{(i)}\big)^\top \widehat{\bm{\Sigma}}_{\widehat{\gamma}}^{-1}\big(\widehat{\bm{\gamma}}- \widehat{\bm{\gamma}}_{(i)} \big)\bigg], \quad i = 1,\ldots,n.
$$

The matrix $\widehat{\bm{\Sigma}}_{^{\widehat \theta}}$ can be approximated by $-\ddot{\bm \ell}^{-1}$. In addition, if we use a first order approximation of the type $ \widehat{\bm{\theta}}- \widehat{\bm{\theta}}_{^{(i)}} \approx \ddot{\bm \ell}_{^{(i)}}^{-1}\,\dot{\bm \ell}_{^{(i)}}$, we obtain
\begin{equation*} \label{GCD}
\textrm{GCD}_i(\bm{\theta}) \approx \frac{1}{q}\left(\dot{\bm \ell}_{(i)}^\top\, \ddot{\bm \ell}_{(i)}^{-1}  (-\ddot{\bm \ell})\, \ddot{\bm \ell}_{(i)}^{-1}\, \dot{\bm \ell}_{(i)}  \right), \quad i = 1,\ldots,n,
\end{equation*}
where $\dot{\bm \ell}_{(i)}$ and $\ddot{\bm \ell}_{(i)}$ are the score vector and Hessian matrix, respectively, without considering the case $i$, evaluated at $\bm{\theta} = \widehat{\bm{\theta}}$.\\

We implement the relative change (RC) to check the impact on the estimated components of the model for the detected influential cases. This measure is defined by computing the estimates removing influential cases and re-estimating the
parameters as well as their corresponding standard errors (s.e) through the expressions
\begin{equation*}
    \mbox{RC}_{\theta_{(i)}}=\left|\frac{\widehat{\theta}_t-\widehat{\theta}_{t(i)}}{\widehat{\theta}_t}\right|\times 100\% \qquad \mbox{and} \qquad \mbox{RC}_{\mbox{se}(\widehat{\theta}_{t(i)})}=\left|\frac{\mbox{se}(\widehat{\theta}_t)-\mbox{se}(\widehat{\theta}_{t(i)})}{\mbox{se}(\widehat{\theta}_t)}\right|\times 100\%,
\end{equation*}
where $\widehat{\theta}_{t(i)}$ and se$(\widehat{\theta}_{t(i)})$ denotes the MLE of $\theta_t$ and its corresponding s.e.

\subsection{Quantile residuals}

Perform a residual diagnostic is crucial to validate a model applied to a data set. Given the simplicity of the c.d.f for the \textbf{RIRON}$_\tau $-F, an ad hoc residual is given by the randomized quantile residuals (rQR) presented in \cite{Dunn1996}. In our case, such residuals are given by
\begin{equation*}
\text{rQR}_i=\Phi^{-1}\left\{F\left[\frac{1}{\lambda}\left(\sqrt{\frac{t_i}{\beta_{\tau i}}}-\sqrt{\frac{\beta_{\tau i}}{t_i}}\right)\right]^{\alpha_\tau}\right\}, \qquad i=1,\ldots,n.
\end{equation*}
If the fitted model is appropriated for the data, $\text{rQR}_1, \ldots, \text{rQR}_n$ should be a random sample for the standard distribution, which can be verified using different normality tests: Kolmogorov-Smirnov (KS), Anderson-Darling (AD), Shapiro-Wilks (SW), Cram\'er-Von Mises (CVM), among others. See \cite{Yap2011} for a discussion about those tests.

\subsection{Computational Aspects} 

To compute the maximum likelihood estimation for $\vb*\theta$, the log-likelihood function defined in \eqref{eq:loglik} must be maximized. We do not obtain an elegant ``closed-form'' solution, but \eqref{eq:loglik} can be maximized using iterative procedures such as the Newton-Raphson, BFGS, BHHH and SANN methods. 

We developed an \verb!R! package called \verb!IRON! that  provides a set of tools for fit and diagnostics of the quantile regression model based in the \textbf{IRON} distribution. For example, the function \verb!quant_reg()! is used to fit quantile regression models, specified by giving a symbolic description of the linear predictor and a description of the kernel. The current version is stored 
on \verb!GitHub! and can be downloaded using
\begin{lstlisting}[language=R]
  devtools::install_github("santosneto/IRON")
\end{lstlisting}

\section{Numerical studies}\label{sec:4}

In this section, we would like to demonstrate the 
performance of the maximum likelihood estimators 
under two scenarios. In our first scenario, we
consider a real data set, while, in the second
scenario, we use artificial data sets.

\subsection{Scenario \#1}

A subset of the data considered here were previously analysed by \cite{https://doi.org/10.1002/asmb.2556}. 
The full data consist of  Chilean House Hold Income for the year 2016. It can be obtained by the National Statistics Institute of Chile, which are available at \url{http://www.
ine.cl/estadisticas/ingresos-y-gastos/esi/base-de-datos}. Based in the application presented in \cite{https://doi.org/10.1002/asmb.2556},
we consider the following variable: household income ($T$), the total income due to salaries ($X_1$), the total income due to independent work ($X_2$)
and the total income due to retirements ($X_3$). In each iteration (1000 replicates), we consider 100 observations selected (randomly) from the full data set and 
we fit the following model
\[
\beta_i = \gamma_1 + \gamma_2 x_{1i} + \gamma_3 x_{2i} + \gamma_4 x_{3i}, \quad i = 1, \ldots, 100,
\]
where $T_i \sim $\textbf{RIRON}$_\tau$-F$(\lambda, \beta_i, \xi; k(\cdot))$ and $\tau$ is fixed.
During the simulations, we consider two cases: (i) we can force the same sample in each iteration and we fit one model for each value of $\tau$ and family (more realistic);
(ii) for each value of $\tau$ (separately) we fit the model considering the families \textbf{N}, $\mathbf{t}$, \textbf{L} and \textbf{EP}. In both cases, our interest is to verify the percentage that each model is selected considering the AIC. The fit was realized by function \verb!quant_reg()! of the \verb!R! package \verb!IRON!. 

Table~\ref{tab:2} we present the percentage that each model is selected considering the AIC - Case I. Here we should analyze the results and check which model is chosen more often considering the different values of $\tau$ and families. The model chosen most often is potentially the best fit. From the results obtained, we can conclude that for $\tau = 0.1, 0.2, 0.3$ and $0.4$, the model with kernel \textbf{EP} presented the best fit for all samples. For $\tau = 0.5$, the model with kernel \textbf{EP} was selected on 53.2\% of the samples. The model with kernel \textbf{t} was selected for most samples when considering $\tau = 0.6$ and $0.7$. Finally, for $\tau = 0.8$ and $0.9$, the model that presented the best performance was the model with kernel \textbf{N}.
%
\begin{table}[ht]
\centering
\caption{Percentage that each model is selected considering the AIC - Case I.}
\begin{tabular}{ccccc} 
  \toprule
\multirow{2}*{$\tau$} & \multicolumn{4}{c}{Kernel}\\ \cmidrule{2-5}
 & Normal & Student'$t$ & Logistic & Exponential Power \\ 
  \midrule
0.1  &  0.0 & 0.0  & 0.0 & \textbf{100.0}\\
0.2  &  0.0 &0.0  & 0.0 &\textbf{100.0}\\
0.3  &  0.0 &0.0 & 0.0 &\textbf{100.0}\\
0.4  &  0.0 &0.0 & 0.0 &\textbf{100.0}\\
0.5 & 0.6 & 35.6 & 10.6 & \textbf{53.2} \\ 
 0.6 & 2.8 & \textbf{69.0} & 28.2 & 0.0 \\ 
 0.7 & 11.7 & \textbf{45.5} & 42.8 & 0.0 \\ 
 0.8 & \textbf{37.1} & 33.3 & 29.6 & 0.0 \\ 
 0.9 & \textbf{70.6} & 25.0 & 4.4 & 0.0 \\ 
\bottomrule
\end{tabular}
\label{tab:2}
\end{table}

Table \ref{tab:3} we present the percentage of times that each model is selected considering the AIC - Case II. For each value of $\tau$, we check which family has the highest percentage of choice, i.e, the best model for a specific value of $\tau$.  Thus, with the results obtained we have to $\tau = 0.6$ and $0.9$ the model with kernel $\mathbf{t}$ is the selected.  While considering $\tau = 0.1, 0.2, 0.3, 0.4$ and $0.5$ we select the model with kernel \textbf{EP}. Finally, note that for $\tau = 0.7$ and $0.8$ the model with kernel \textbf{L} is the model selected.

\begin{table}[ht]
\centering
\caption{Percentage that each model is selected considering the AIC - Case II.}
\begin{tabular}{ccccc} 
  \toprule
\multirow{2}*{$\tau$} & \multicolumn{4}{c}{Kernel}\\ \cmidrule{2-5}
 & Normal & Student'$t$ & Logistic & Exponential Power \\ 
  \midrule
0.1  &  0.0 & 0.0  & 0.0 & \textbf{100.0}\\
0.2  &  0.0 &0.0  & 0.0 &\textbf{100.0}\\
0.3  &  0.0 &0.0 & 0.0 &\textbf{100.0}\\
0.4  &  0.0 &0.0 & 0.0 &\textbf{100.0}\\
0.5  &  0.1 &29.8  & 13.1 &\textbf{57.0}\\
0.6  &  1.6 &\textbf{62.6}  &35.8 &0.0\\
0.7  & 5.2 & 41.5 & \textbf{53.3}& 0.0\\
0.8  & 20.4 &  37.3 & \textbf{42.3}& 0.0\\
0.9  & 40.1 & \textbf{57.3}  & 2.6 &0.0\\
\bottomrule
\end{tabular}
\label{tab:3}
\end{table}

\subsection{Scenario \#2}

This study employs a Monte Carlo simulation
to evaluate the performance of four kernels 
under different size sample . In this scenario,  the aim is to analyse the return of the maximum likelihood estimates of the parameters of the models. We evaluated
their performances by measuring their relative bias (RB) and root mean squared error (RMSE) based on 5000 replications of the model
\[
\beta_i = 0.5 + 1.5 x_{i1} - 0.5 x_{i2}, \quad i = 1, \ldots, n_{\text{obs}}, \quad (n_{\text{obs}} \in \{30, 100, 600\}),
\]
where $x_{i1}$ and $x_{i2}$ are distributed uniformly 
on the interval (0,1). The values generated this variables are kept fixed on each replication.  We assume that $T_i \sim \textbf{RIRON}(2.0, \beta_i, \xi; k(\cdot))$, considering the kernels: \textbf{N}, $\mathbf{t}$, \textbf{L} and \textbf{EP}. For the kernels $\mathbf{t}$ and \textbf{EP} we have that $\xi = 4.0$ and for other kernels $\xi$ is not defined. A pseudo-random number generator
was used to produce independent random variables $T_i$. We use the function \verb!riron()!, which generates random variables from the IRON distribution, and the fit was realized by function \verb!quant_reg()! both of the \verb!R! package \verb!IRON!. 

Table~\ref{tab:tab4} we present the RB and RMSE estimated under different kernels. The result in Table~\ref{tab:tab4} are now discussed for each distribution.
\begin{enumerate}
    \item \textbf{Normal}: The RB of the MLE's of $\lambda$, $\gamma_1$, $\gamma_2$ and $\gamma_3$ are negative. For the MLE of $\gamma_1$, only for $n_{\text{obs}} = 600$ the RB is negative. The reduction of the RB's of the estimators were: 96\% ($\widehat{\lambda}$), 93\% ($\widehat{\gamma}_1$), 62\% ($\widehat{\gamma}_2$) and 50\% ($\widehat{\gamma}_3$). As we can see, in Table~\ref{tab:tab4}, the greatest reduction 
    occur for $\widehat{\lambda}$. In the case of RMSE we can observe that all the estimates present reduction with the increase of the sample size. We can highlight the estimator $\widehat{\gamma}_2$, which presented a reduction of 81\%. 
    \item \textbf{Student't}: For the model with kernel Student't note that the MLE, which presented greater reduction of RB was $\widehat{\xi}$ with a reduction of 99.8\%. Now, for the RMSE we have that the MLE, which presented greater reduction was also $\widehat{\xi}$ with 99.6\%.
    \item \textbf{Logistic}: Analysing the results obtained, we can see that the MLE which presented the lower reduction of RB with the increase of the sample size was $\widehat{\gamma}_3$ . This MLE presented a reduction of 50.6\%. This represents a value 41,8\% lower than the average reduction of the others MLE's. Besides, for RMSE we have that the worst performing MLE is $\widehat{\gamma}_1$ with reduction of 78.9\%.   
    \item \textbf{Exponential Power}: Finally, we discuss the results for the kernel \textbf{EP}. For this kernel note that the MLE of $\xi$ is quite biased for small sample. For example, for $n = 30$ the RB is 120.294. However, with the growth of the sample, this bias falls sharply (for $n = 600$ the RB is 0.042). The same situation is observed for the RMSE.
\end{enumerate}

\begin{table}[H]
    \centering
    \caption{RB and RMSE estimated for $\tau = 0.5$, $\lambda = 2.0$ and $\xi = 4.0$ under different kernels.}
    \begin{tabular}{c ccccc c ccccc}\toprule
    \multirow{2}*{$n$}     & \multicolumn{5}{c}{RB} &&\multicolumn{5}{c}{RMSE} \\ \cmidrule{2-6} \cmidrule{8-12} 
         &  $\widehat{\lambda}$ & $\widehat{\xi}$ & $\widehat{\gamma}_1$ & $\widehat{\gamma}_2$ & $\widehat{\gamma}_3$ & &  $\widehat{\lambda}$ & $\widehat{\xi}$ & $\widehat{\gamma}_1$ & $\widehat{\gamma}_2$ & $\widehat{\gamma}_3$ \\ \cmidrule{1-12}
         \multicolumn{12}{c}{Normal}\\ \cmidrule{2-12} 
        30 &  -0.077 & $\times$ & 0.101 &-0.016& -0.056 && 0.303& $\times$ & 0.386 & 0.728&0.517  \\
        100 & -0.021&$\times$&0.030&-0.011&-0.022 &&0.147&$\times$ &0.211 &0.352 & 0.240  \\
        600 &-0.003&$\times$ &-0.007 & -0.006 & -0.028 &&0.059&$\times$ & 0.092 &0.141 & 0.103  \\ 
    \midrule
    \multicolumn{12}{c}{Student' t}\\ \cmidrule{2-12}
        30 & -0.029 & 34.515 &0.273 &0.080&0.081 && 0.563 & 250.135& 0.628  &     1.155  &     0.806 \\
        100 & 0.022 & 4.510 &0.056&-0.010&-0.024 && 0.282  &85.532 &0.296    &   0.497 &      0.333  \\
        600 & 0.005 & 0.075 &$<$0.001&-0.009&-0.028 && 0.106  &0.945 &0.126   &    0.192    &   0.140\\
    \midrule
    \multicolumn{12}{c}{Logistic}\\ \cmidrule{2-12}
        30 & -0.069 &$\times$ & 0.220  &-0.031 &-0.081 &&0.348  &$\times$ &0.588 &1.052 & 0.786 \\
        100 &-0.018  &$\times$ &  0.065 & -0.022&       -0.028 && 0.173  &$\times$ &0.297 & 0.487       &0.340 \\
        600 &-0.003  &$\times$ &  -0.006& -0.010      &-0.040 && 0.070 &$\times$ &0.124 & 0.192 & 0.140 \\
    \midrule
    \multicolumn{12}{c}{Exponential Power}\\ \cmidrule{2-12}
        30 & -0.080   & 120.294 &      0.057&       0.015 &     -0.029 && 0.346&    653.225&       0.281&       0.485&       0.396  \\
        100 &-0.003  &    0.377 &      0.009&      -0.001&      -0.007 && 0.156&      3.556&       0.125&       0.210 &      0.144 \\
        600 &0.001     & 0.042&      -0.003 &     -0.002    &  -0.011 && 0.062&      0.553&       0.052&       0.081&       0.058 \\    
    \bottomrule     
    \end{tabular}
    \label{tab:tab4}
\end{table}

\section{Application}\label{sec:5}

In this Section we reanalized a data set related to the household income in Chile illustrated in \cite{https://doi.org/10.1002/asmb.2556}
based on the \textbf{RBSQ} model. 

\subsection{Household income in Chile}

This data set corresponds to chilean house-hold income in the year 2016 (the last study avaliable up to this moment) collected by the National Institute of Statistics, Chile, which are available at https://www.ine.cl/estadisticas/sociales/ingresos-y-gastos/encuesta-de-presupuestos-familiares. \cite{https://doi.org/10.1002/asmb.2556} consider a subsample of size $n = 100$ cases randomly selected from the full data set
and for comparative purposes we considered the same cases. The response variable is the household income in thousands of Chilean pesos ($Y$), whereas the covariates to be considered in the analysis are the total income due to salaries ($X_1$), the total income due to independent work ($X_2$) and the total income due to retirements ($X_3$). A descriptive analysis of the variables is presented in Table \ref{tab:desc1}. 

\begin{table}[H]
    \centering
    \caption{Descriptive statistics for income data (in thousands of Chilean pesos).}
     \begin{tabular}{crrrrrrrrr}\toprule
Variable & Mean & Median & SD & CV & CS & CK & Min & Max & $n$ \\
\midrule
$Y$   & 938.14 & 698.78 & 837.52 & 0.89 & 2.45 & 11.03 & 70.00 & 5369.95 & 100\\
$X_1$ & 401.26 & 254.42 & 547.91 & 1.37 & 2.32 & 10.21 &  0.00 & 3231.38 & 100\\ 
$X_2$ & 172.46 &   0.00 & 467.60 & 2.71 & 3.96 & 19.90 &  0.00 & 3005.01 & 100\\
$X_3$ & 88.63  &   0.00 & 212.56 & 2.40 & 3.53 & 16.81 &  0.00 & 1299.32 & 100\\
\bottomrule
 \end{tabular}
 \label{tab:desc1}
\end{table}

\begin{figure}[H]
\begin{subfigure}[b]{0.5\textwidth}
\centering
\includegraphics[width=\textwidth]{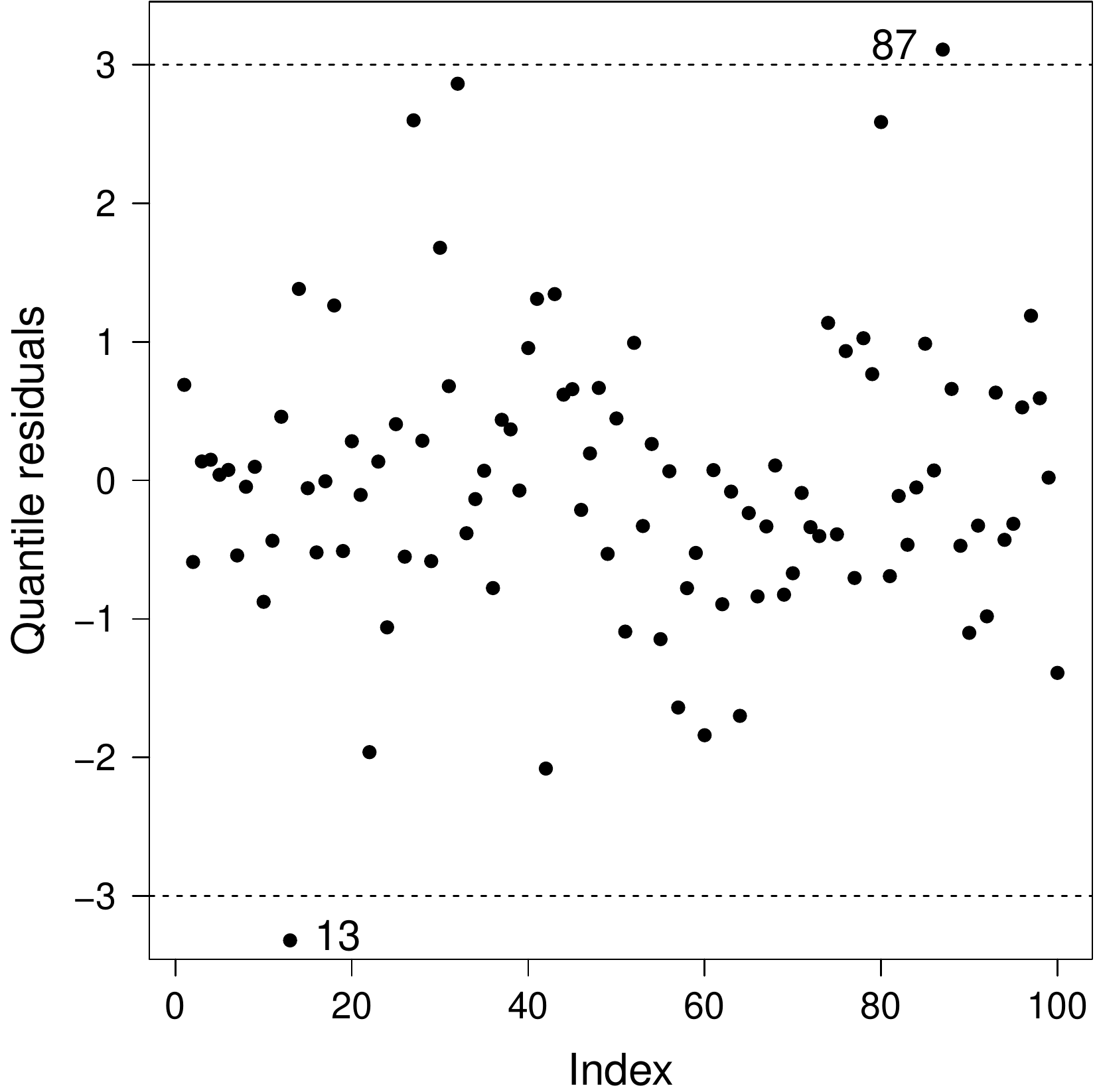}
 \caption{\textbf{N}}
\end{subfigure}
\hfill
\begin{subfigure}[b]{0.5\textwidth}
\centering
\includegraphics[width=\textwidth]{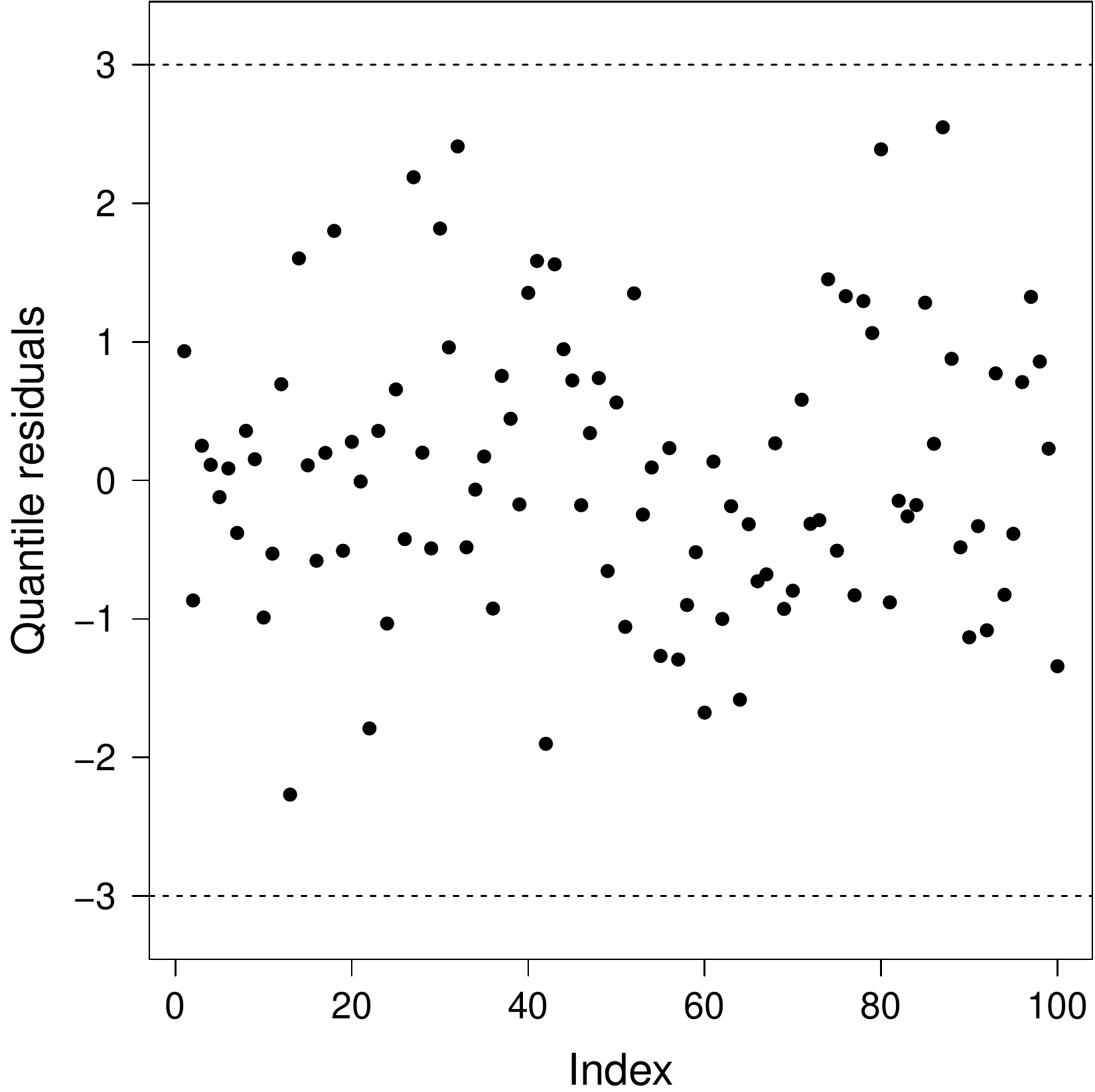}
 \caption{\textbf{t}}
\end{subfigure}
\hfill
\begin{subfigure}[b]{0.5\textwidth}
\centering
\includegraphics[width=\textwidth]{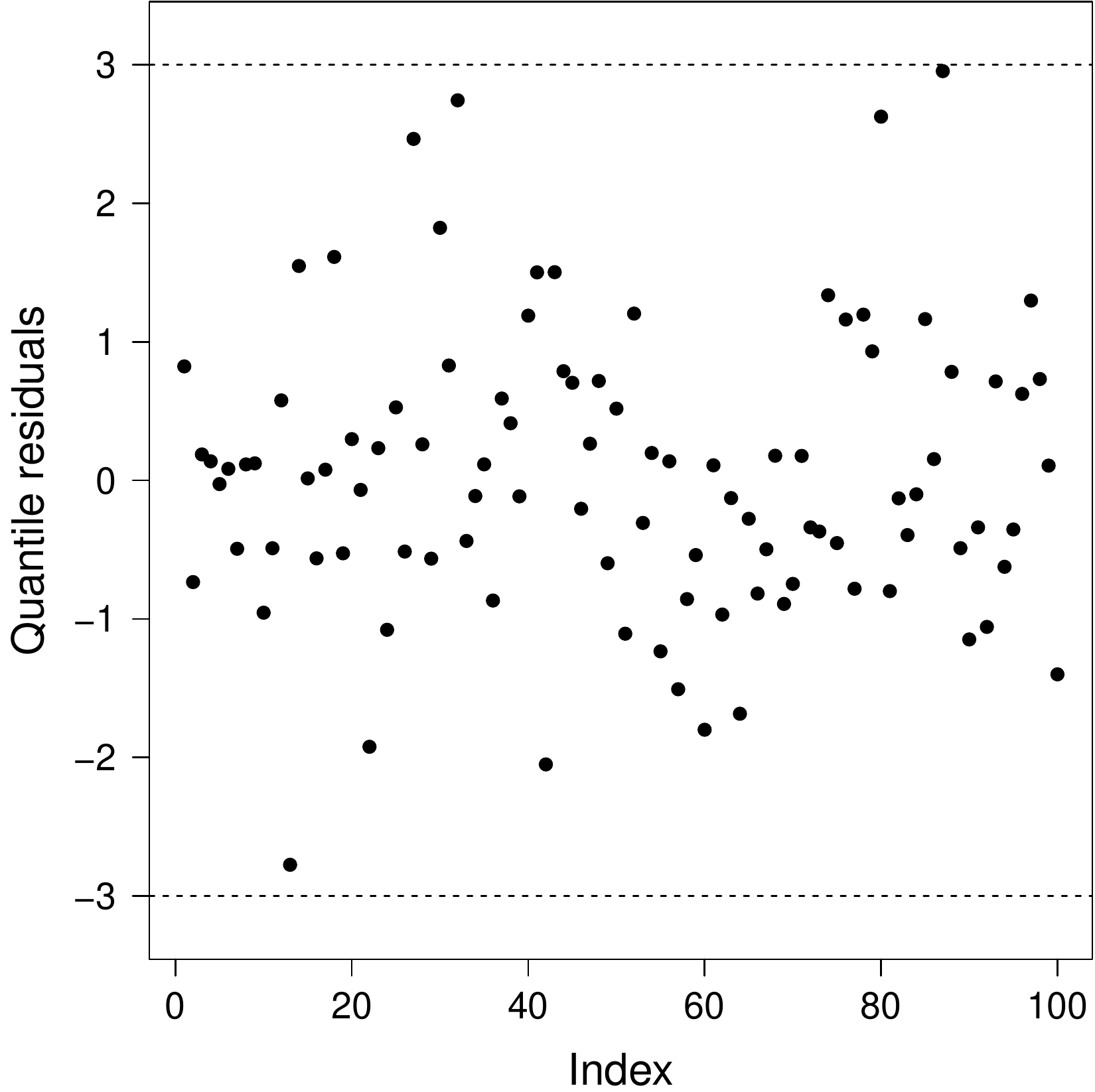}
\caption{\textbf{L}}
\end{subfigure}
\hfill
\begin{subfigure}[b]{0.5\textwidth}
\centering
\includegraphics[width=\textwidth]{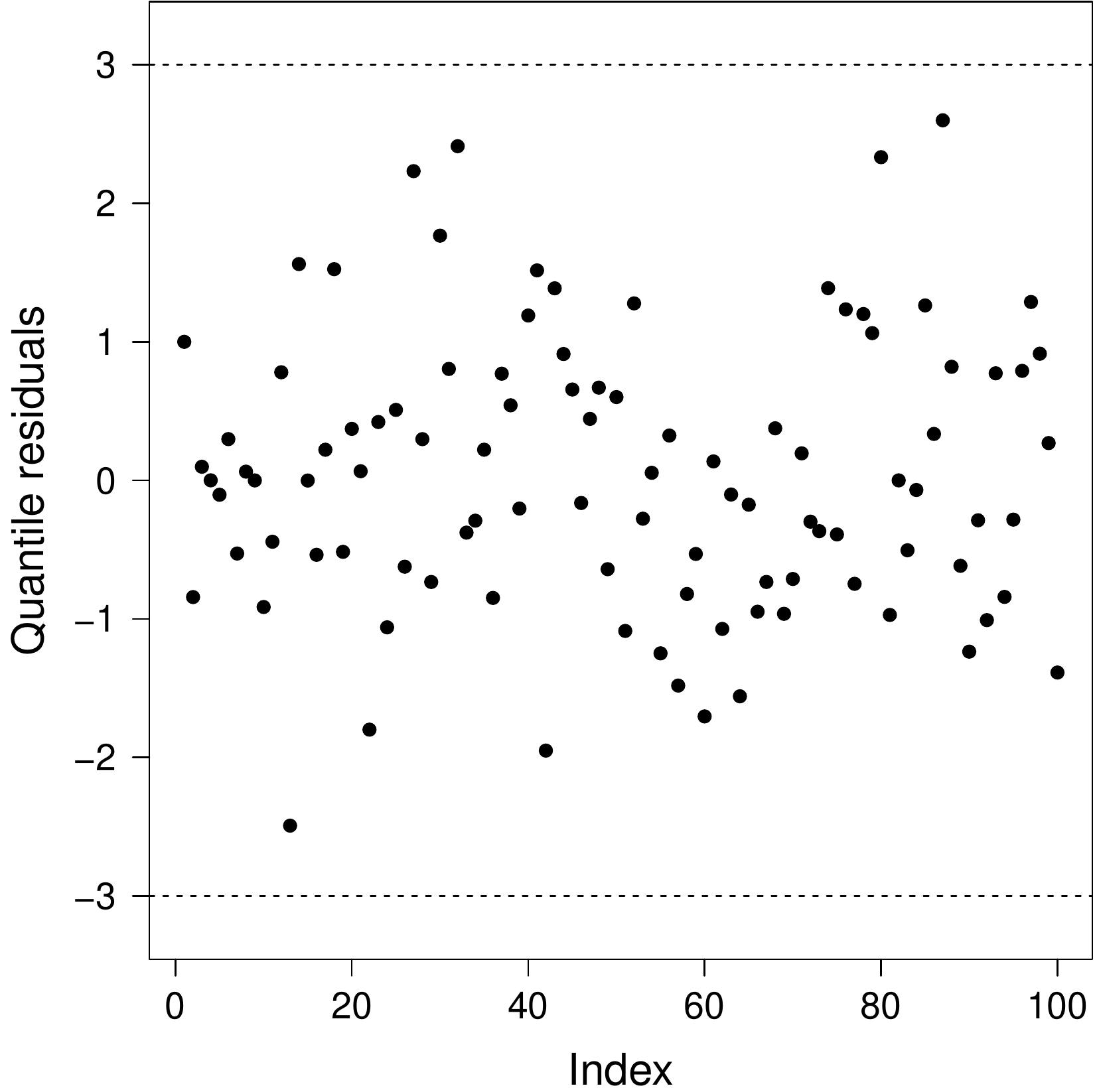}
 \caption{\textbf{EP}}
\end{subfigure}
\caption{rQRs for different quantile regression model for $\tau=0.5$ in the \textbf{RIRON} class: \textbf{N}, \textbf{t}, \textbf{L} and \textbf{EP}.}
\label{graf:QR}
\end{figure}

\begin{figure}[H]
\begin{subfigure}[b]{0.5\textwidth}
\centering
\includegraphics[width=\textwidth]{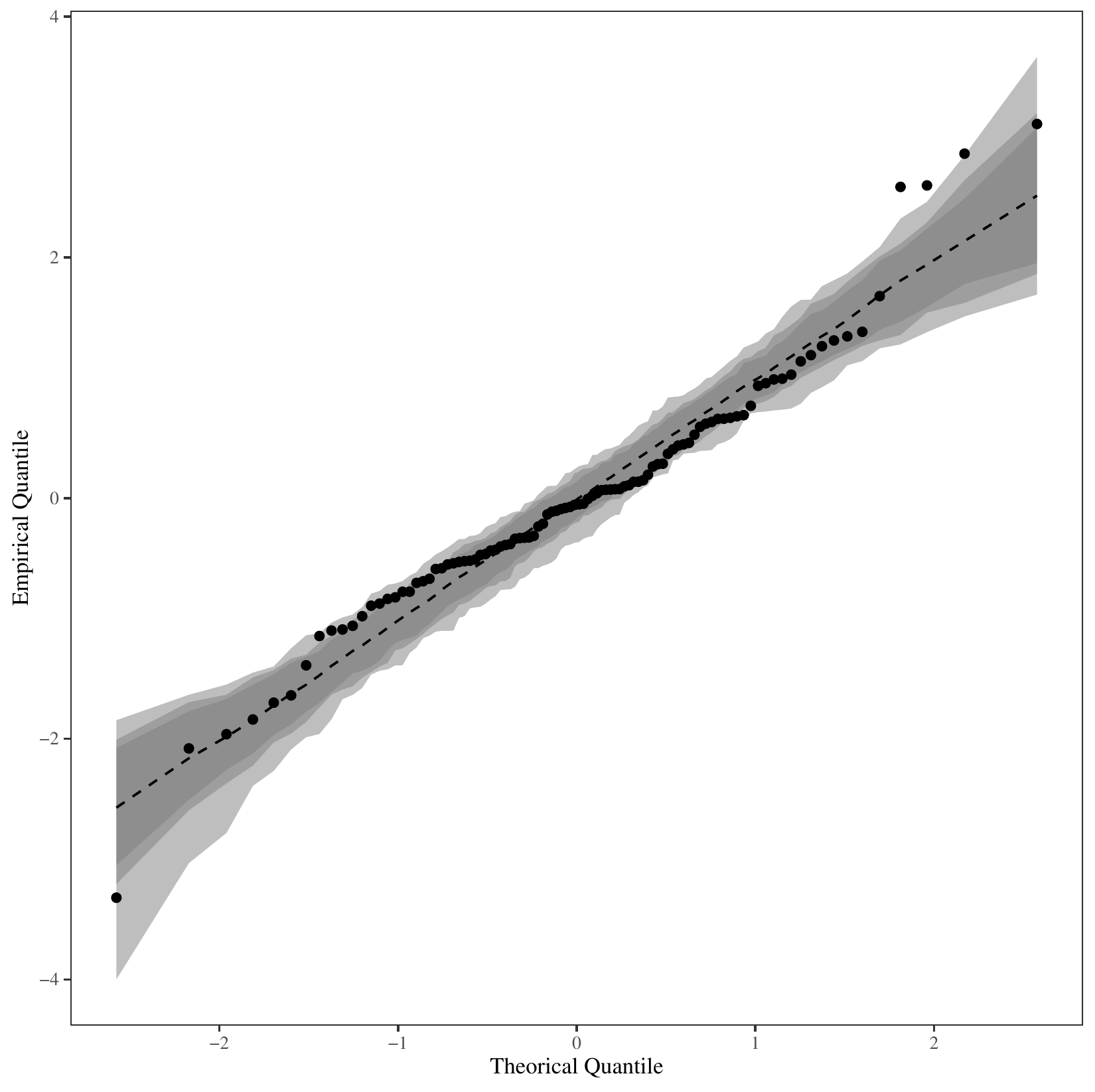}
\caption{\textbf{N}}
\end{subfigure}
\hfill
\begin{subfigure}[b]{0.5\textwidth}
\centering
\includegraphics[width=\textwidth]{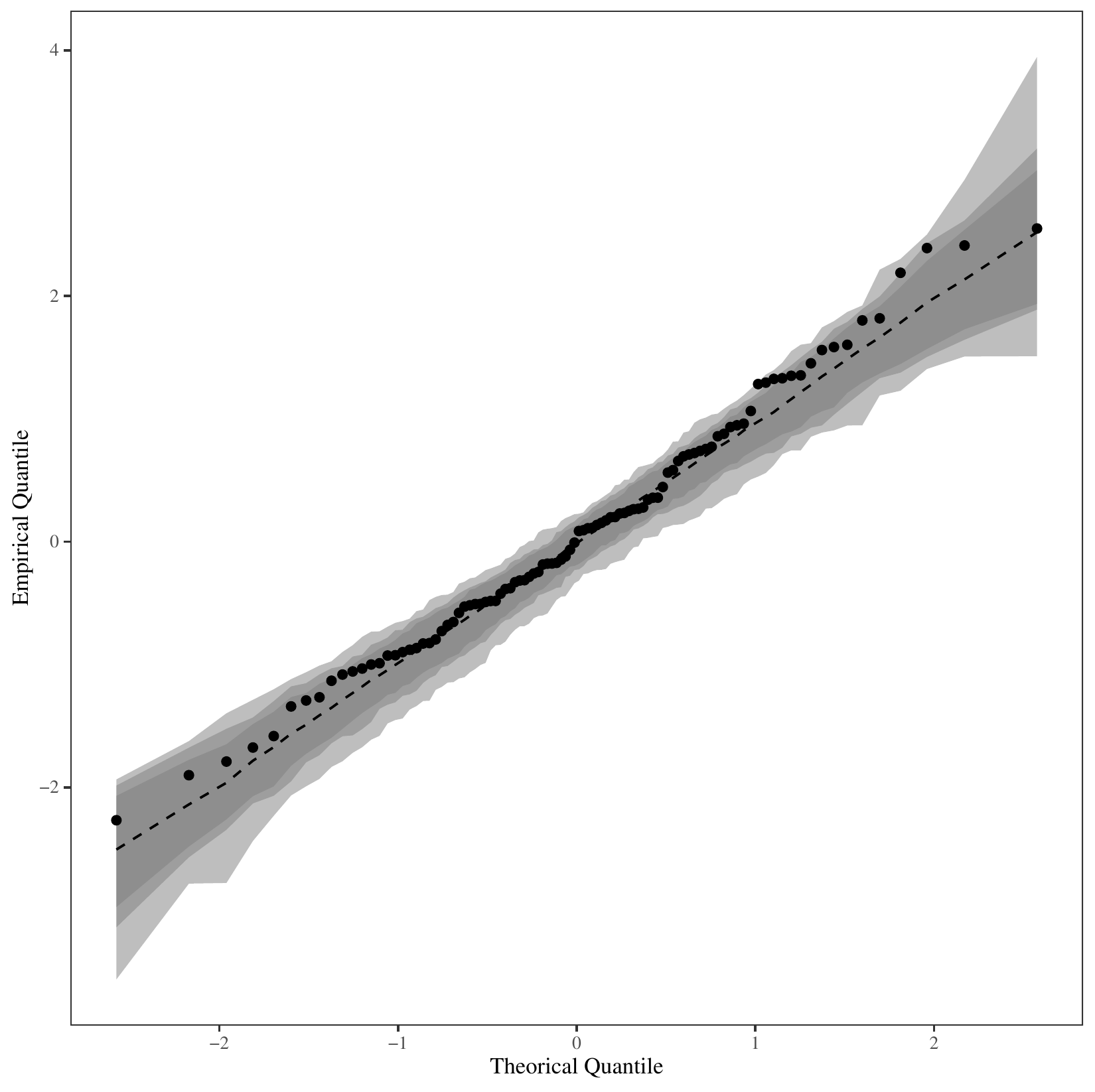}
\caption{\textbf{t}}
\end{subfigure}
\hfill
\begin{subfigure}[b]{0.5\textwidth}
\centering
\includegraphics[width=\textwidth]{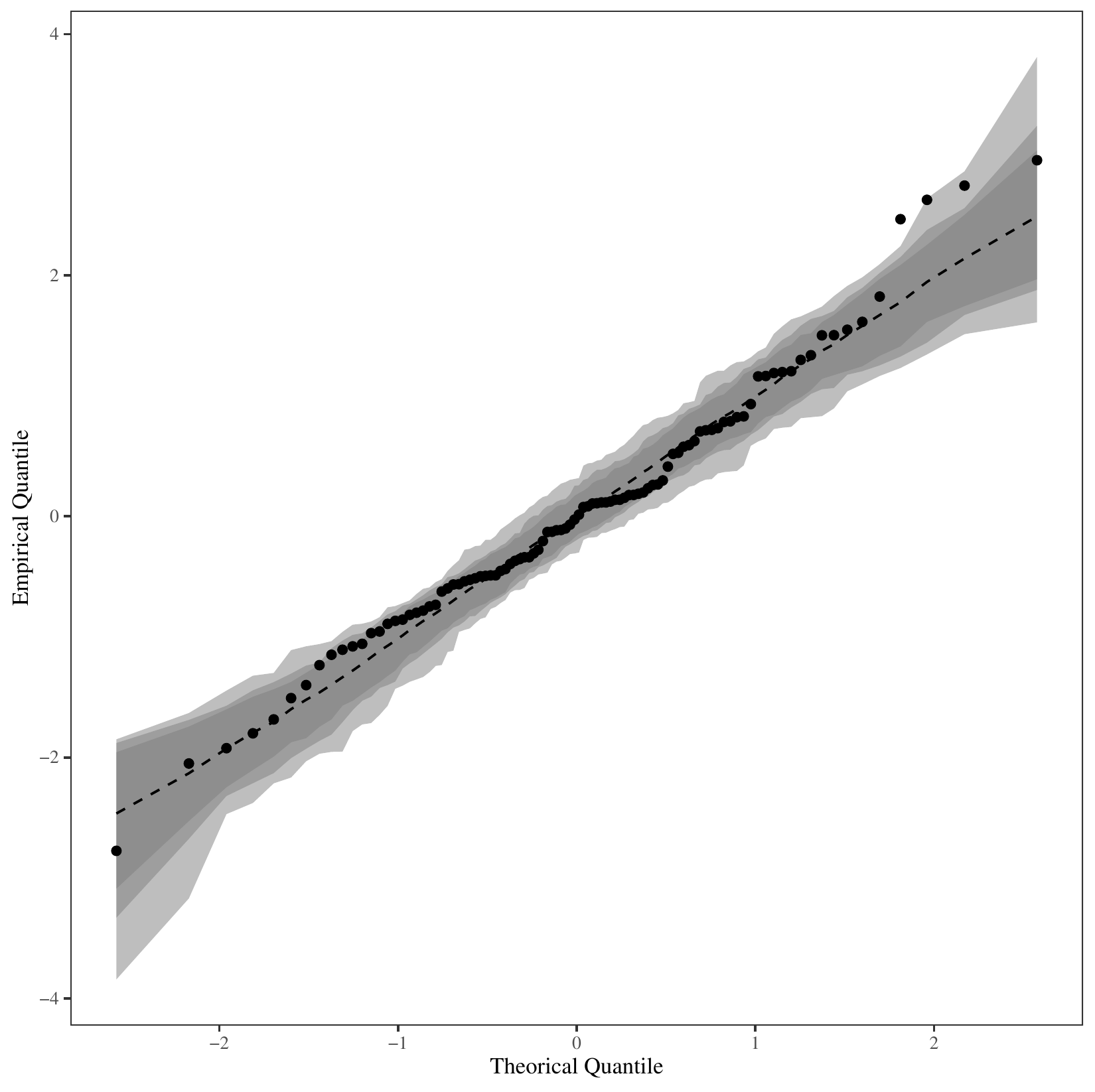}
\caption{\textbf{L}}
\end{subfigure}
\hfill
\begin{subfigure}[b]{0.5\textwidth}
\centering
\includegraphics[width=\textwidth]{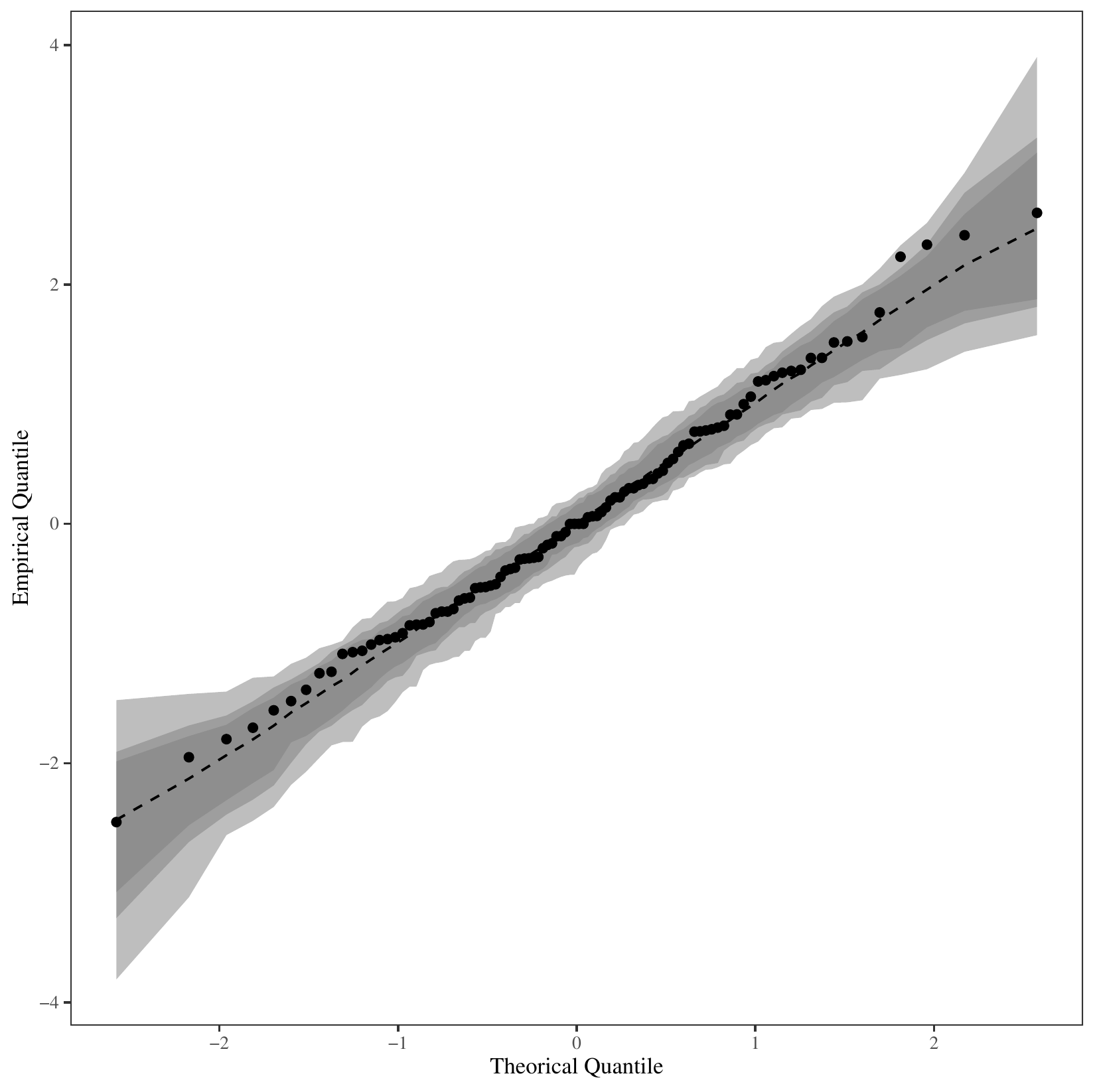}
 \caption{\textbf{EP}}
\end{subfigure}
\caption{Envelopes for the QRs for different quantile regression model for $\tau=0.5$ in the \textbf{RIRON} class: \textbf{N}, \textbf{t}, \textbf{L} and \textbf{EP}.}
\label{graf:env}
\end{figure}

\begin{figure}[H]
\begin{subfigure}[b]{0.5\textwidth}
\centering
\includegraphics[width=\textwidth]{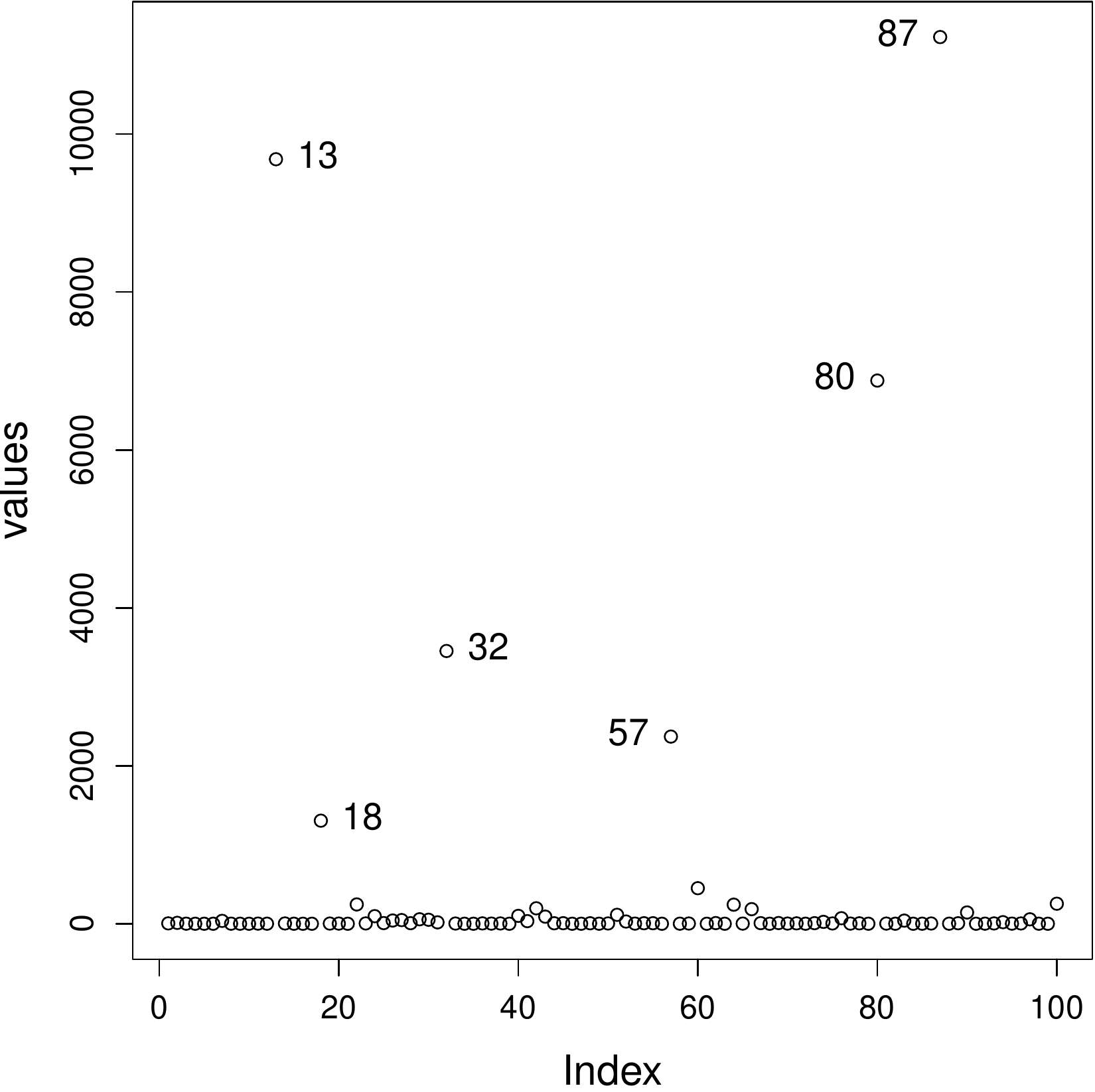}
\caption{\textbf{N}}
\end{subfigure}
\hfill
\begin{subfigure}[b]{0.5\textwidth}
\centering
\includegraphics[width=\textwidth]{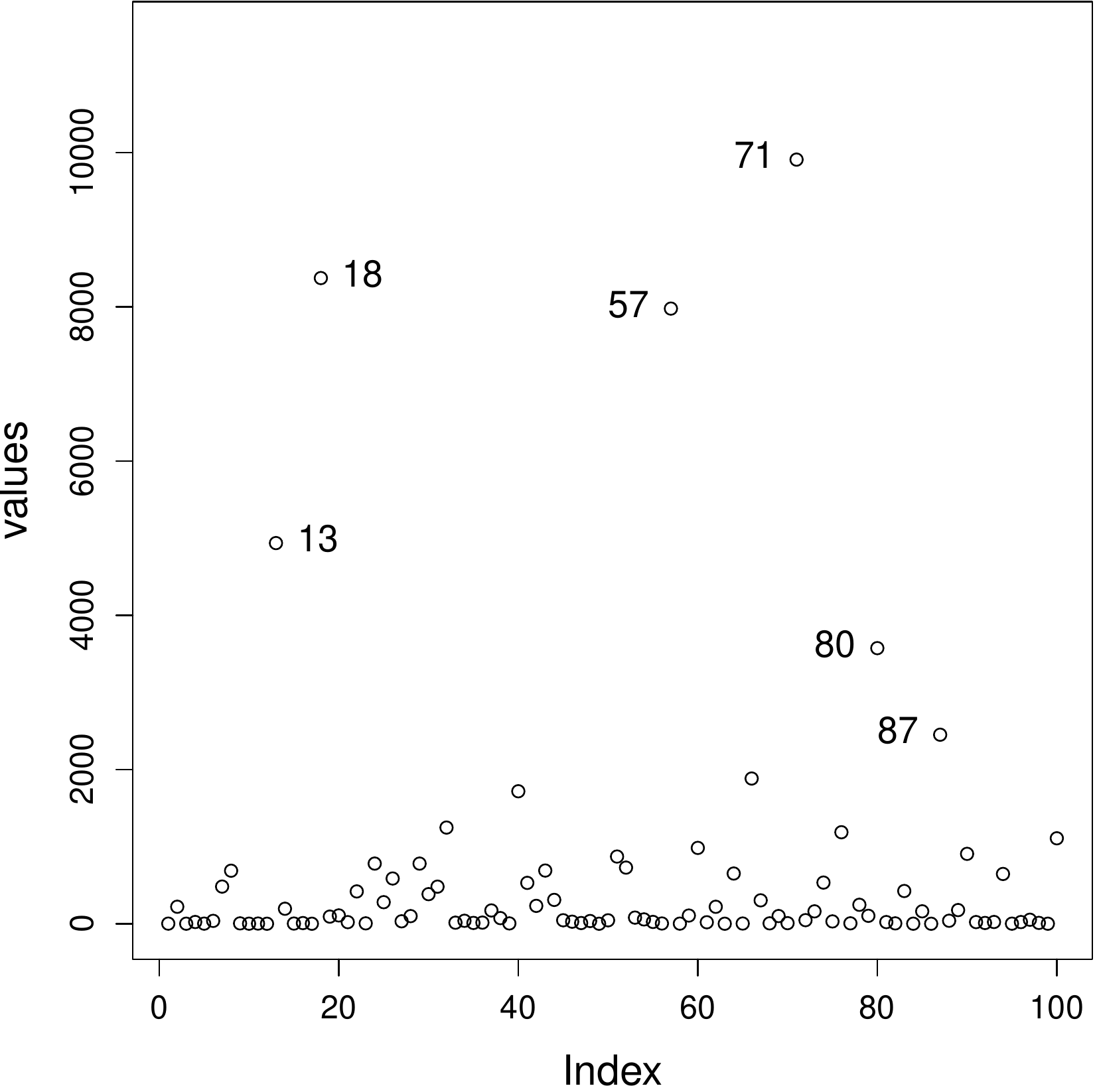}
\caption{\textbf{t}}
\end{subfigure}
\hfill
\begin{subfigure}[b]{0.5\textwidth}
\centering
\includegraphics[width=\textwidth]{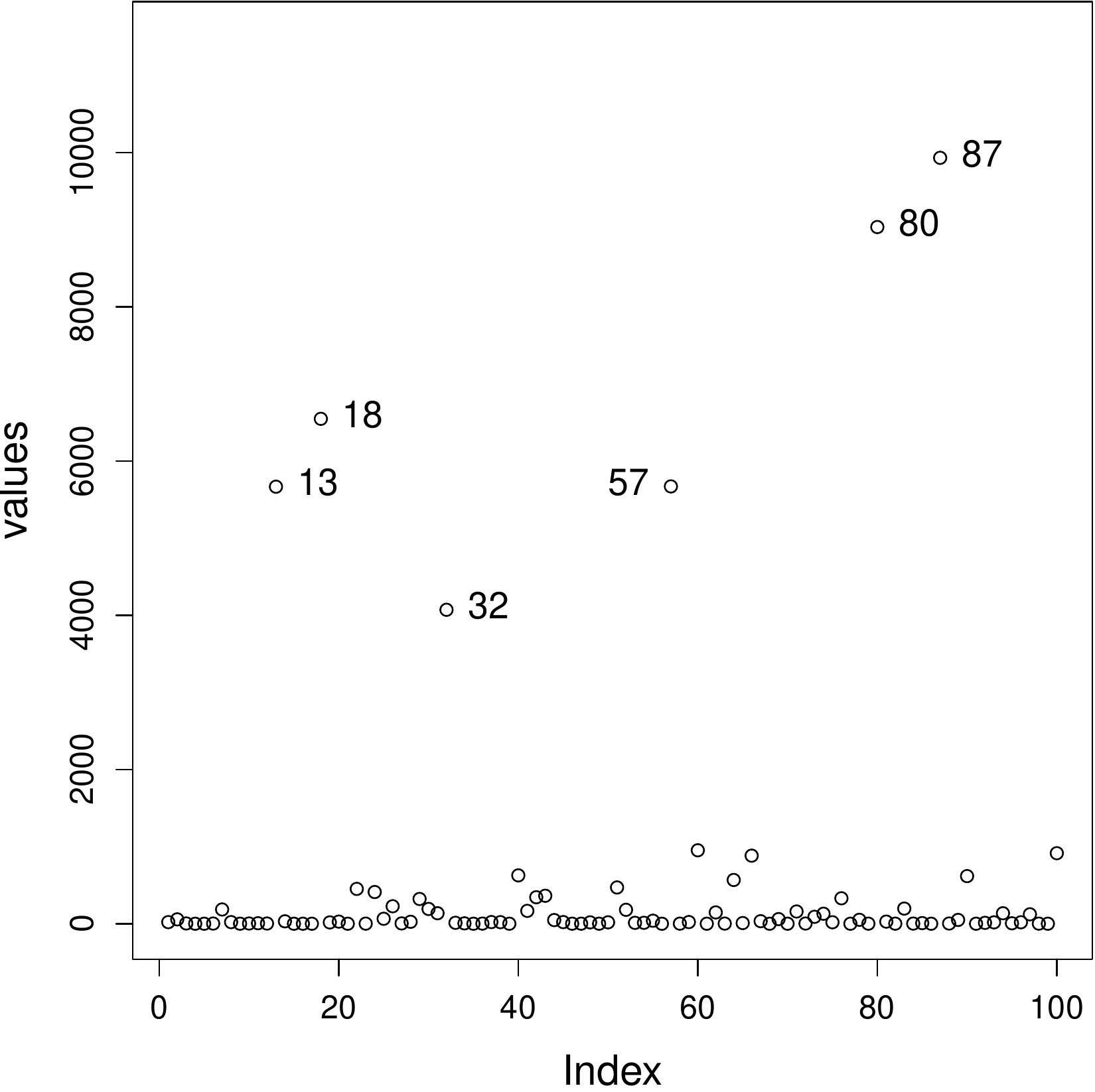} 
\caption{\textbf{L}}
\end{subfigure}
\hfill
\begin{subfigure}[b]{0.5\textwidth}
\centering
\includegraphics[width=\textwidth]{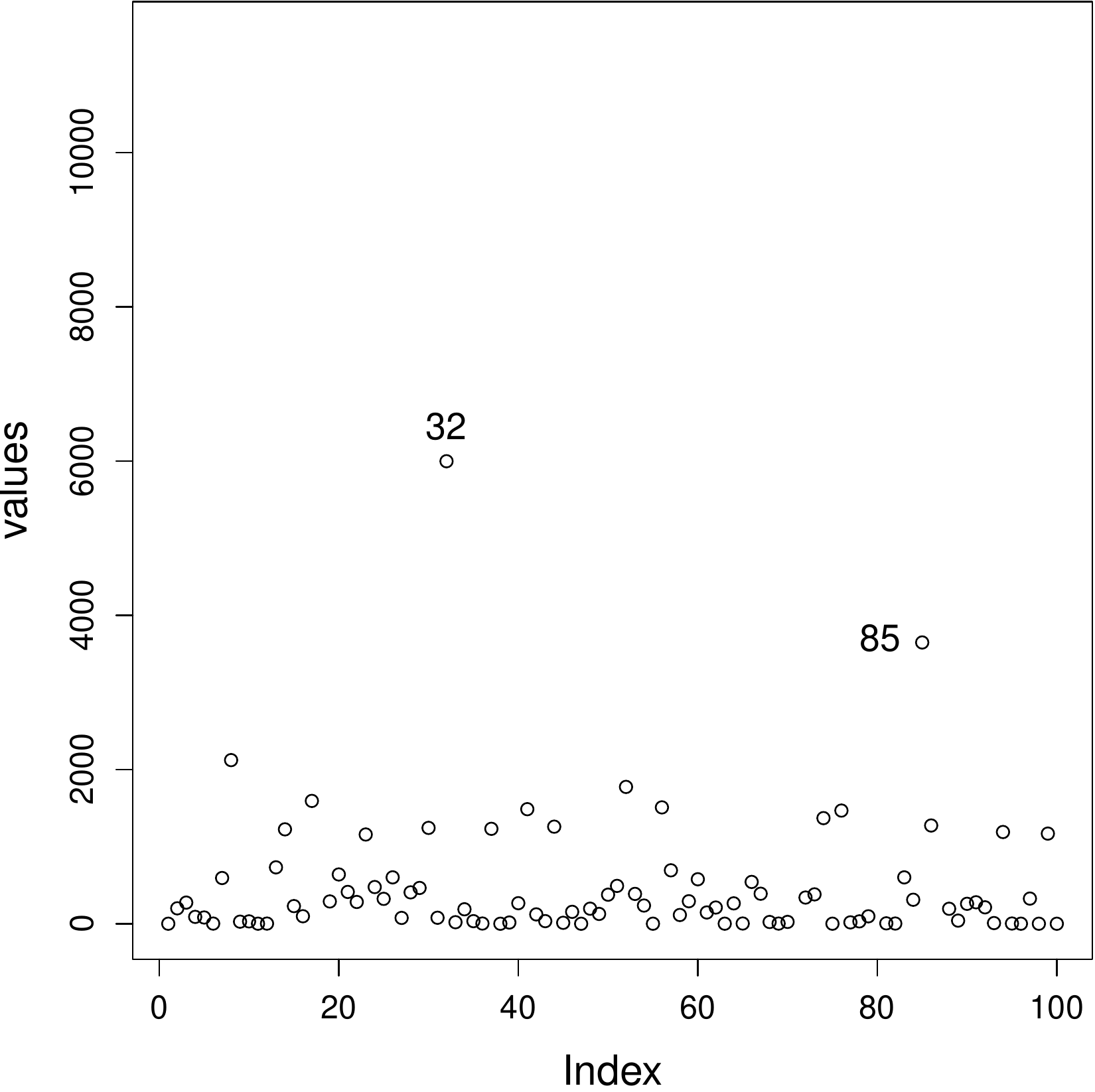}
 \caption{\textbf{EP}}
\end{subfigure}
\caption{Cook's distance for different quantile regression model for $\tau=0.5$ in the \textbf{RIRON} class: \textbf{N}, \textbf{t}, \textbf{L} and \textbf{EP}.}
\label{graf:cook}
\end{figure}

We propose to consider that $Y_i\sim $\textbf{RIRON}$\tau$-F$(\beta_i, \lambda)$, where
\begin{equation*}
\beta_{\tau i}=\beta_0 + \beta_1 X_1 + \beta_2 X_2 +\beta_3 X_3, \quad i=1,\ldots,n.
\end{equation*}
We considered $\tau=0.5$ (the median) in order to compare our proposal with the model in \cite{https://doi.org/10.1002/asmb.2556}. Additionally, we also considered $\tau=0.4$ because in Chile the most of the social benefits are given to families that belong to the lowest 40\% of the country's income (see http://www.registrosocial.gob.cl/beneficios-sociales). For this data set and considering the log, identity and sqrt link, the better results are given for the identity model. Therefore, we focused in this link. AIC and BIC criteria are presented in Table \ref{tab:est1}. Note that the \textbf{RIRON-EP} model attached the minimum AIC and BIC criteria among the fitted models for this data set in both, $\tau=0.4$ and $\tau=0.5$. We also highlight that for the median ($\tau=0.5$), the \textbf{RIRON-N} model coincides with the \textbf{RBSQ} model and for $\tau=0.4$ the AIC and BIC are 1395.7 and 1408.7, respectively, higher than the obtained for the models in Table \ref{tab:est1}. We present the QRs in Figure \ref{graf:QR} the QRs. Note that for the \textbf{N} case, there are two points out of range $(-3,3)$ for those residuals, whereas for the rest of models there are no points outside of such range. Figure \ref{graf:cook} also shows the Cook's distance for the fitted models, where the \textbf{RIRON-EP} model has less possible influent observations (2 versus 6 for the rest of the models), suggesting that this model is more robust in comparison with its competitors. Finally, Figure \ref{graf:env} shows the envelope for the QRs, where also the \textbf{RIRON-EP} shows a better fit for those residuals in relation to the normal distribution. The three graphical tool suggest than the \textbf{RIRON-EP} model provides a better fit for this data set than the rest of the models.\\
To illustrate the difference among the estimation of different models, we consider a family with 3 people contributing to the household income: one who works for the minimum wage in Chile (after social security and health discounts approximately 265 thousands of Chilean pesos) and two retirees receiving the minimum value of a pension (approximately 160 thousands of Chilean pesos), without independent work. Under this characteristics, the \textbf{RIRON-N} model estimate that the most vulnerable 40\% of the Chilean population receives at most 750.264 thousands of Chilean pesos, whereas under the \textbf{RIRON-EP} model such estimate is at most 808.199 thousands of Chilean pesos. In other words, the \textbf{RIRON-N} model underestimate the household income in the value of approximately one basic food basket, a measure used in Chile representing the minimum threshold of requirements to a person obtain 2,000 calories per day in a month in Chile (see \url{http://observatorio.ministeriodesarrollosocial.gob.cl/storage/docs/cba/nueva_serie/2021/Valor_CBA_y_LPs_21.01.pdf}).

Figure \ref{graf:env} shows the envelopes for the rQR for the four models in the \textbf{RIRON} class for $\tau=0.5$. The minimum $p$-value for the KS, AD, SW and CVM tests are 0.3265, 0.0917 and 0.7542 for the \textbf{IRON-t}, \textbf{IRON-L} and \textbf{IRON-EP} models, respectively, suggesting that the three mentioned models are reasonable for this data set. However, the AD, SW and CVM have a related $p$-value of 0.0064, 0.0052 and 0.0131 for the \textbf{IRON-N} model, respectively. This result suggests that the \textbf{IRON-N} (which matchs with the \textbf{RBSQ} model for the median case) is not appropriate for this data set.

\begin{table}[H]
    \centering
    \caption{Descriptive statistics for income data (in thousands of Chilean pesos).}
\begin{tabular}{c c r r r r r r r r}
\toprule
\multirow{3}*{$\tau$}& \multirow{3}*{Parameter} & \multicolumn{8}{c}{Kernel}\\
\cmidrule{3-10}
           &            &  \multicolumn{2}{c}{Normal} &  \multicolumn{2}{c}{Student't} &  \multicolumn{ 2}{c}{Logistic } & \multicolumn{ 2}{c}{ Exponential Power} \\
\cmidrule{3-10}
        &  &   estimate &       s.e. &   estimate &       s.e. &   estimate &       s.e. &   estimate &       s.e. \\
\midrule
\multirow{6}*{$0.4$} &      $\beta_0$ &    177.729 &     10.695 &    138.737 &     13.086 &    158.542 &     12.328 &    157.814 &      6.062 \\

           &      $\beta_1$ &      0.959 &      0.108 &      1.017 &      0.064 &      0.995 &      0.081 &      0.969 &      0.024\\

           &      $\beta_2$ &      1.021 &      0.194 &      1.182 &      0.113 &      1.105 &      0.153 &      1.253 &      0.052\\

           &      $\beta_3$ &      0.995 &      0.144 &      1.109 &      0.121 &      1.044 &      0.134 &      1.230 &      0.100\\

           &      $\lambda$ &      0.394 &      0.021 &      0.249 &      0.038 &      0.207 &      0.015 &      0.151 &      0.065\\

           &   $\xi$/$\kappa$ &  \multicolumn{ 2}{c}{-} &      2.557 &      1.085 &  \multicolumn{ 2}{c}{-} &      0.743 &      0.148\\
\cmidrule{2-10}
           &        AIC & \multicolumn{ 2}{c}{1395.1} & \multicolumn{ 2}{c}{1382.8} & \multicolumn{ 2}{c}{1385.8} & \multicolumn{ 2}{c}{{\bf 1358.0}} \\

           &        BIC & \multicolumn{ 2}{c}{1408.1} & \multicolumn{ 2}{c}{1398.5} & \multicolumn{ 2}{c}{1398.8} & \multicolumn{ 2}{c}{{\bf 1373.6}} \\
\midrule

\multirow{6}*{$0.5$}        &      $\beta_0$ &    198.148 &     11.727 &    163.689 &     15.638 &    181.628 &     14.152 &    183.442 &     15.122 \\

           &      $\beta_1$ &      1.044 &      0.119 &      1.057 &      0.071 &      1.050 &      0.086 &      0.999 &      0.058 \\

           &      $\beta_2$ &      1.109 &      0.218 &      1.216 &      0.131 &      1.161 &      0.169 &      1.223 &      0.112  \\

           &      $\beta_3$ &      1.086 &      0.158 &      1.159 &      0.136 &      1.118 &      0.148 &      1.118 &      0.124  \\

           &      $\lambda$ &      0.365 &      0.019 &      0.243 &      0.036 &      0.191 &      0.014 &      0.262 &      0.080  \\

           &   $\xi$/$\kappa$ &  \multicolumn{ 2}{c}{-} &      3.046 &      1.382 &  \multicolumn{ 2}{c}{-} &      1.001 &      0.210  \\
\cmidrule{2-10}
           &        AIC & \multicolumn{ 2}{c}{1395.7} & \multicolumn{ 2}{c}{1386.0} & \multicolumn{ 2}{c}{1387.1} & \multicolumn{ 2}{c}{{\bf 1385.4}} \\

           &        BIC & \multicolumn{ 2}{c}{1408.7} & \multicolumn{ 2}{c}{1401.6} & \multicolumn{ 2}{c}{1400.1} & \multicolumn{ 2}{c}{{\bf 1401.0}}  \\
\bottomrule
\end{tabular}  
 \label{tab:est1}
\end{table}

On the other hand, the robustness of the \textbf{IRON-EP} model is illustrated through the GDC in Figure \ref{graf:cook}, because the \textbf{IRON-EP} model has only two potential influent observations versus six observations for the rest of fitted models. As illustrated in Table \ref{tab:est.dropped}, the main differences in the estimation with and without the highlighted observations are produced for the SE terms for the \textbf{IRON-EP} model. However, such differences no modify the significance of any parameter.

\begin{table}[H]
    \centering
    \caption{RCs (in \%) in ML estimates and their corresponding se's for the indicated parameter and dropped
cases and respective $p$-values for the \textbf{IRON-PE} model in household income in Chile.}
\begin{tabular}{ccrrrrrr}
\toprule
\multirow{2}*{Dropped Cases}&\multirow{2}*{Component} & \multicolumn{6}{c}{Parameter}\\ 
\cmidrule{3-8}
 &   &      $\beta_0$ &      $\beta_1$ &      $\beta_2$ &      $\beta_3$ &     $\lambda$ &      $\kappa$ \\
\midrule

 \multirow{3}*{32} &         RC$_{\widehat{\theta}_{t(i)}}$ &     6.84 &     0.22 &     1.79 &     5.58 &     1.14 &     3.35 \\

           &         RC$_{\mbox{se}(\widehat{\theta}_{t(i)})}$ &     4.98 &     1.92 &     2.29 &     0.73 &     1.12 &     2.25 \\

           &    $p$-value &    $<$0.0001 &    $<$0.0001 &    $<$0.0001 &    $<$0.0001 &     0.0008 &    $<$0.0001 \\
\midrule

   \multirow{3}*{85} &         RC$_{\widehat{\theta}_{t(i)}}$ &     5.33 &     3.74 &     1.67 &     4.35 &    17.58 &    11.49 \\

           &         RC$_{\mbox{se}(\widehat{\theta}_{t(i)})}$ &     9.76 &    20.45 &    35.25 &    17.14 &     3.25 &    10.79 \\

           &    $p$-value &    $<$0.0001 &    $<$0.0001 &    $<$0.0001 &    $<$0.0001 &     0.0053 &    $<$0.0001 \\
\midrule

  \multirow{3}*{\{32, 85\}} &         RC$_{\widehat{\theta}_{t(i)}}$ &    12.02 &     3.20 &     2.45 &     9.27 &    20.76 &    11.67 \\

           &         RC$_{\mbox{se}(\widehat{\theta}_{t(i)})}$ &    15.23 &    29.63 &    24.84 &    35.44 &    10.85 &    15.73 \\

           &    $p$-value &    $<$0.0001 &    $<$0.0001 &    $<$0.0001 &    $<$0.0001 &     0.0036 &    $<$0.0001 \\
\bottomrule
\end{tabular}  
 \label{tab:est.dropped}
\end{table}

\section{Discussion and Conclusion}\label{sec:6}

Quantile regression and the Birnbaum-Saunders distribution have been widely used in many fields. Based on the family of distributions called standard $\alpha$-exponentiated Birnbaum-Saunders and standardized symmetrical distributions, we built a family of distributions called \textbf{IRON}. The manuscript presents, $\textbf{RIRON}_\tau \mathrm{-F}(\beta_\tau,\lambda)$, a quantile regression framework that allows to model different levels of flatness. The approach combines the idea of propose a quantile regression model considering the \textbf{IRON} distribution. This model generalizes the model proposed by \cite{https://doi.org/10.1002/asmb.2556} and makes modeling flexible. Experimental study with one real-world dataset demonstrated the good performance of the presented method. In particular, the new approach with kernel \textbf{EP} presented promissory results.

\section*{Acknowledgment}
"Research carried out using the computational resources of the Center for Mathematical Sciences Applied to Industry (CeMEAI) funded by FAPESP (grant 2013/07375-0)."

\end{document}